\documentclass[runningheads]{cl2emult}

\usepackage{makeidx}  
\usepackage{graphicx} 
\usepackage{subeqnar} 
\usepackage{multicol} 
\usepackage{cropmark} 
\usepackage{eso}      
\makeindex            

\begin{document}
\title*{The Evolution of Ellipticals, Spirals and Irregulars: 
Overcoming Selection Bias}

\toctitle{The Evolution of Ellipticals, Spirals and Irregulars}

%
%
\titlerunning{Evolution and Selection Bias}
%
\author{Simon Driver}
\authorrunning{Simon Driver}
%
%
\institute{University of St Andrews, St Andrews, KY16 9DW, SCOTLAND}

\maketitle              

\begin{abstract}
The Hubble Deep Fields represent our best opportunity for probing
galaxy evolution over a substantive look-back time. However as with
any dataset the HDFs are prone to selection biases. These biases
are extremely severe beyond $z \sim 1.25$ such that a meaningful 
interpretation of generic galaxy evolution is not possible. We can however
extract well defined volume-limited samples at $z < 1$. The data are
entirely consistent with passive/null-evolution for ellipticals, spirals and
irregulars however this concluion is tempered by small number statistics.
Alas stringent constraints on galaxy evolution await an 
order of magnitude increase in the number of HDFs.
\end{abstract}

\section{Introduction}
The two Hubble Deep Fields (HDF) provide our deepest optical insight into 
the near and far universe. Each contains approximately 750 galaxies
for which magnitudes, colours, photometric redshifts and morphologies
have been determined \cite{D98}. The data have been used by many 
groups to construct a picture of the assembly and evolution of 
galaxies. However every dataset, the HDFs included, suffer from
selection biases at some level. Here I hope to briefly highlight the
severity of these selection biases and to provide some suggestions as 
to how to test for, and monitor, these biases. In particular 
I advocate the use of the Bivariate Brightness Distribution (BBD) 
as a means of probing galaxy evolution while also tracking the
biases which plague our ability to decipher deep data.
Finally I put forward my own cautious conclusions as to the evolution
of ellipticals, spirals and irregulars since z=1.

\section{Selection Bias}
\begin{figure}
\hspace{-1.0cm}
\noindent
\includegraphics[width=0.6\textwidth]{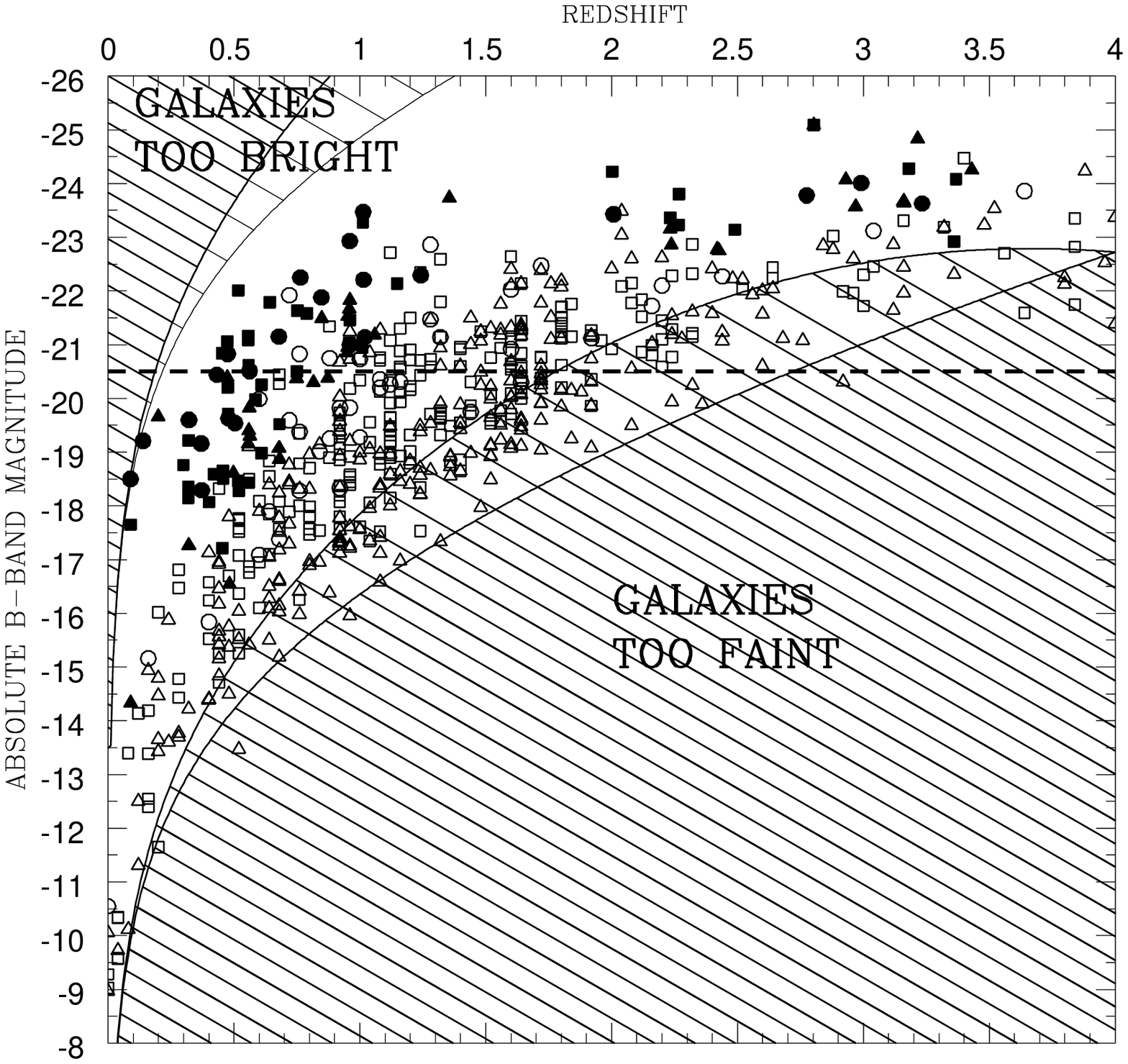}
\hspace{-1.0cm}
\includegraphics[width=0.6\textwidth]{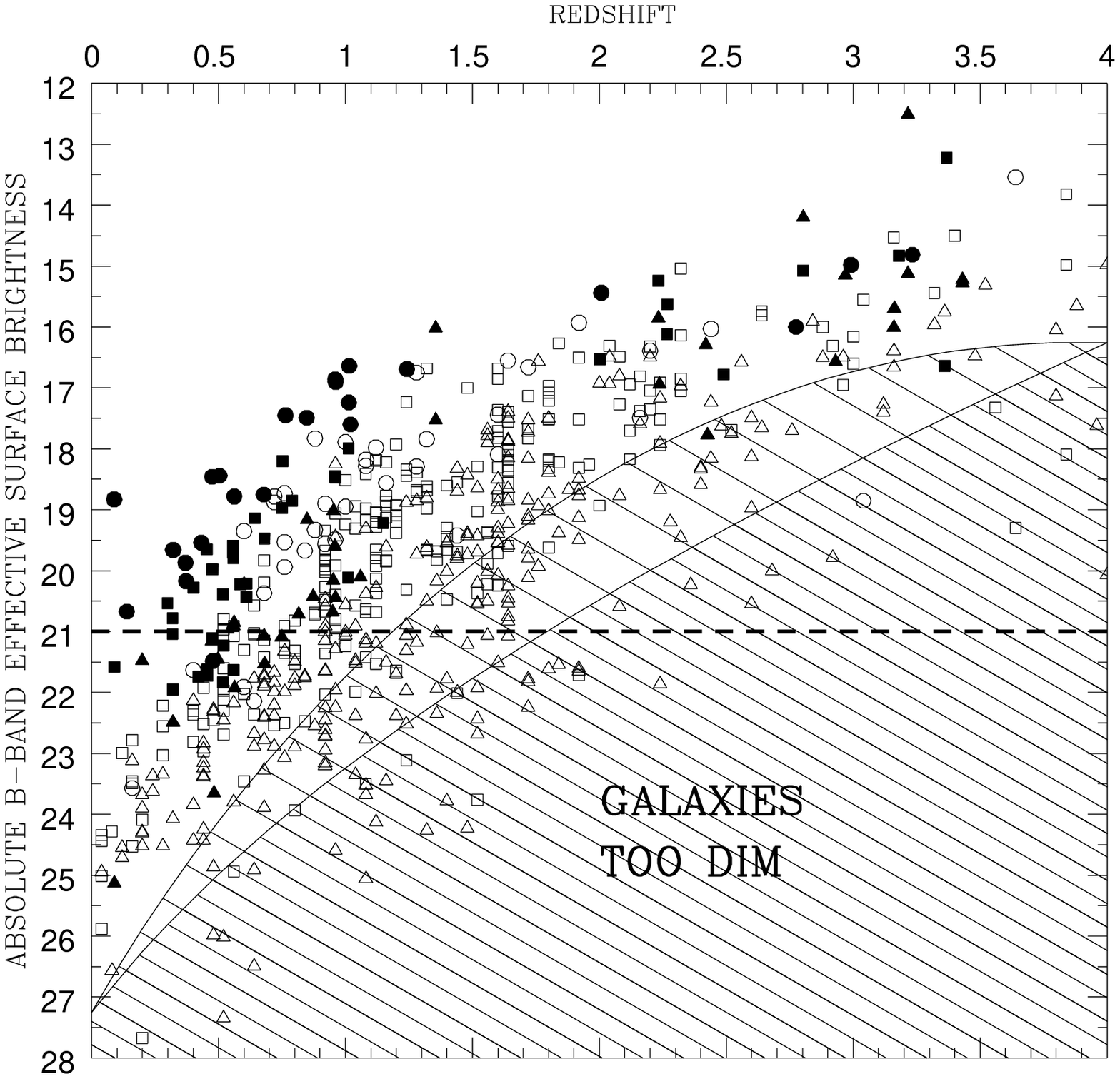}
\caption{These two plots ($M$ v $z$, left) and ($\mu$ v $z$, right)
highlight the severe effect of Malmquist- and Disney- Bias which
impinge define the selection window through which we view the
galaxy population at any redshift. The two selection limits define
the uncertainty due to the range of possible K-corrections. Solid points are
spectrscopic redshifts, circles are ellipticals, squares spirals and
triangles irregulars}
\label{fig1}
\end{figure}

Galaxies exist over a vast dynamic range in both luminosity and
in surface brightness. Any dataset has an apparent magnitude limit, 
a surface brightness detection limit and size limits which together 
create a selection window \cite{D00}.
We view of the galaxy population through this selection window
Different surveys with differing 
limitations will sample different regions of the galaxian parameter-space.
Unless these biases are monitored discrepancies will manifest between 
surveys as are seen for example in measures of the local luminosity function
\cite{CD01}. These concerns become more paramount when one 
compares local and deep data, as such biases will cause discrepancies 
that may erroneously be interpreted as evolution.
Additionally as these selection biases are all distance dependent the 
selection window varies such that comparisons within a single dataset are
also susceptible. The most important of these biases is
Malmquist Bias - i.e., as we go fainter we select
against firstly, low luminosity systems, then normal systems and eventually
giant systems. An analogous bias, Disney Bias \cite{D76}, works in a similar 
way but in terms of a galaxy's absolute surface brightness such that lower
surface brightness systems are preferentially lost first, etc.
Neglecting either Malmquist or Disney Bias leads to misleading
results and as both biases are ever present it is vital that we, as a 
community, demonstrate the robustness of our data to these biases
prior to interpretation. Fig.~\ref{fig1} shows two visualisations of 
the HDF data highlighting the severity of both the Malmquist bias (left)
and the Disney bias (right). The dashed lines shows the location of the
canonical non-evolving zero-redshift $L_{*}$ point.

The conclusion we must draw from Fig.~\ref{fig1} is that the HDF
suffers from severe bias beyond z=1 (for the spectroscopic sample:solid points)
and beyond z=1.25 (for the photometric sample). Any statistical
measurement of luminosity functions, star-formation history, evolution
etc, will therefore be incomplete and increasingly misleading beyond z=1.25.
So, does the mean star-formation of the Universe really peak at z=1.5 
and then turn down ? Fig.~\ref{fig1} convinces me that we simply
don't know until we look into the ``unobserved'' regions.
Looking at Fig. 1 it is also easy to see why one might
be drawn to the conclusion that galaxies were both more luminous and
of higher surface brightness in the past. Again an alternative and
more informed conclusion is that this interpretation {\it may} be
purely due to the shifting selection window - we can not say.

~

\noindent
{\bf The conclusion is that the Hubble Deep Fields lack the breadth
in luminosity and surface brightness beyond z=1.25 to place meaningful
constraints on galaxy evolution.}

\section{The Bivariate Brightness Distribution}
Having concluded that a statistical investigation of galaxy evolution 
from the HDFs is currently impossible beyond z=1.25 this confines us to 
investigation galaxy evolution at $z < 1.25$. This is not to say that
selection biases are not present below $z < 1.25$ but that they have
not yet impinged upon the canonical z=0 $L_{*}$ point. Why is the
$L_{*}$ point so critical ? mainly because it is these galaxies
which dominate the luminosity and mass density of the local universe
\cite{C00}.
To see where and whether they have to evolved we have to be able to
``observe'' any positive-, zero- and negative- evolution about this
point.

Additional selection lines also exist which cannot be shown on 
Fig.~\ref{fig1}. These are the size constraints due to the pixel size 
(limiting the highest measurable surface brightness) and the 
field-of-view size (limiting the lowest surface brightness object 
detectable for any absolute magnitude at some redshift). To show these
additional selection lines we must simultaneously look at the $M$-$\mu$ 
plane (our window into the galaxy population). This plane is known as 
the Bivariate Brightness Distribution and has up until no only be derived 
for the Virgo cluster \cite{FB94} - nevertheless it is a powerful
tool.

\begin{figure}
\centering
\noindent
\includegraphics[width=0.7\textwidth]{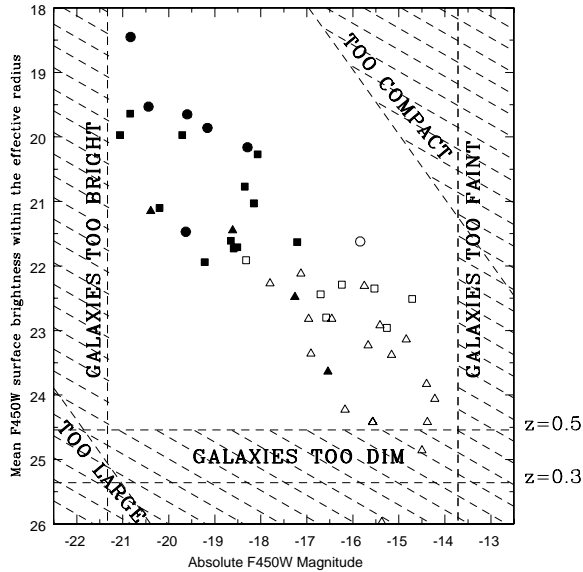}
\caption{The bivariate brightness distribution for a volume limited
(0.3 < z < 0.5) sample showing the various selection lines at work.}
\label{fig3}
\end{figure}

An example BBD, taken from \cite{D00}, is shown as Fig.~\ref{fig3}.
This shows the HDF window into the galaxy population at z=0.3-0.5.
The absolute magnitude, surface brightness and size selection lines
are all shown and the shaded regions indicate areas where the HDF is 
unable to detect galaxies. The selection lines are all redshift dependent 
and control our insight into the galaxy population. The selection lines 
are fairly straight-forward to estimate and are laid out in detail 
in \cite{D00}. Note that those galaxies which appear in the excluded 
region {\it must} have erroneous photometric redshifts as their detection 
{\it is} impossible. As an aside the data show a strong luminosity-surface 
brightness relation and a dearth of luminous low surface brightness
galaxies~\cite{D00}.

~

\noindent
{\bf Using a bivariate brightness distribution we can extract 
volume-limited samples for which the selection biases are well defined.}

\section{The Evolution of Ellipticals, Spirals and Irregulars}
\begin{figure}
\centering
\noindent
\includegraphics[width=1.0\textwidth]{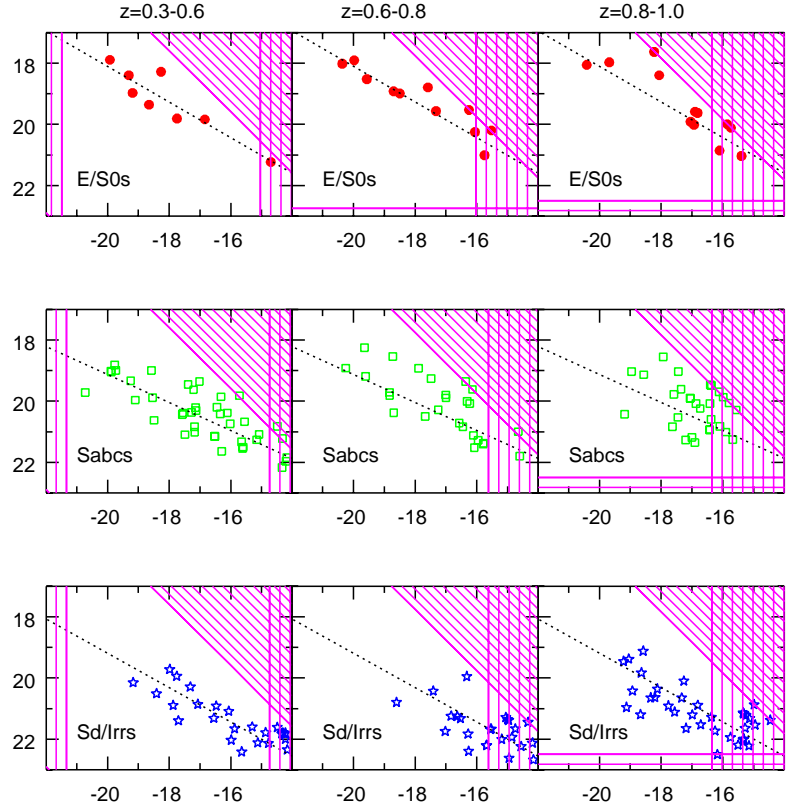}
\caption{The BBD plots for ellipticals
(top), spirals (middle) and irregulars (bottom) for redshift
intervals of 0.3-0.6 (left), 0.6-0.8 (centre), 0.8-1.0 (right)}
\label{fig4}
\end{figure}

Having advocated
a methodology to extract well controlled samples below $z=1.25$,
we are now in a good position to investigate galaxy evolution since $z=1$.
Fig.~\ref{fig4} shows a sequence of BBDs for all the HDF ellipticals, 
spirals and irregulars from z=0 to z=1 for $B<26.5$.
The striped areas show the prohibited regions and galaxies that lie within 
these regions most likely have erroneous photometric redshifts. The
density of galaxies which lie in these excluded regions is of some
concern and highlights the desperate need to spectroscopically confirm
late-types in particular.

Finally then what does Fig.~\ref{fig4} tell us about the evolution of
the three generic galaxy types. The dashed line guides the eye
and are fitted to the z=0.3-0.6 bin for each type. Two conclusions
strike home: (1) the HDF contain insufficient data for a detailed
statistical analysis particularly when one folds in the clustering
uncertainty, and (2), in all cases the 0.3-0.6 fit matches the
0.8-1.0 fit. 

In closing it appears that the current outlook is somewhat
depressing, the HDF is insufficiently deep to probe the evolution of
the generic population beyond $z \sim 1.25$ and the volume insufficient
to probe evolution below $z \sim 1.25$. We desperately need an order of 
magnitude more HDFs.

~

\noindent
{\bf For the moment the Occam's razor solution must surely prevail:
within the limited statistics available the HDF is consitent with
null/passive evolution since $z=1$, beyond this even the HDF data are 
dominated by selection effects.}

\clearpage
\addcontentsline{toc}{section}{Index}
\flushbottom
\printindex

\end{document}